\documentclass[runningheads]{llncs}
\usepackage[T1]{fontenc}
\newcommand{\Figure}[1]{Figure~\ref{fig:#1}}

\newcommand{\Table}[1]{Table~\ref{tab:#1}}

\newcommand{\Equation}[1]{Equation~\ref{eq:#1}}

\newcommand{\Section}[1]{Section~\ref{sec:#1}}

\usepackage{amsmath}
\usepackage{bbold}
\renewcommand\vec{\mathbf}
\newcommand{\mat}{\mathbf}

\DeclareMathOperator*{\argmin}{argmin}

\usepackage{enumitem}
\setlist[itemize]{noitemsep, topsep=0pt}
\setlist[enumerate]{nosep}

\usepackage{caption}
\usepackage{subcaption}

\usepackage{multirow}
\usepackage{graphicx}

\usepackage{sidecap}

\usepackage{float}

\usepackage[flushleft]{threeparttable}

\usepackage{amssymb}
\usepackage{bm}

\newcommand{\nosemic}{\renewcommand{\@endalgocfline}{\relax}}
\newcommand{\dosemic}{\renewcommand{\@endalgocfline}{\algocf@endline}}
\let\oldnl\nl
\newcommand{\nonl}{\renewcommand{\nl}{\let\nl\oldnl}}
\usepackage{afterpage} 

\usepackage{lipsum}  

\usepackage{wrapfig}

\usepackage{lmodern}
\usepackage{enumitem}
\usepackage{hhline}
\usepackage{hyperref}
\usepackage{graphicx}
\begin{document}
\title{Deep Multi-contrast Cardiac MRI Reconstruction via vSHARP with Auxiliary Refinement Network}
\titlerunning{Deep MCCMRI Reconstruction via vSHARP with ARN}
%
\author{George Yiasemis \inst{1,2},
Nikita Moriakov\inst{1,2},
Jan-Jakob Sonke\inst{1,2},
Jonas Teuwen\inst{1,2,3}\thanks{Corresponding author \\ \textit{Email Address:} \email{j.teuwen@nki.nl}}
}

\authorrunning{G. Yiasemis et al.}
%
\institute{Netherlands Cancer Institute, Amsterdam, Netherlands \\
\and
University of Amsterdam,  Amsterdam, Netherlands\\
\and
Radboud University Medical Center, Nijmegen, Netherlands\\
}



%
\maketitle              
\begin{abstract}

Cardiac MRI (CMRI) is a cornerstone imaging modality that provides in-depth insights into cardiac structure and function. Multi-contrast CMRI (MCCMRI), which acquires sequences with varying contrast weightings, significantly enhances diagnostic capabilities by capturing a wide range of cardiac tissue characteristics. However, MCCMRI is often constrained by lengthy acquisition times and susceptibility to motion artifacts. To mitigate these challenges, accelerated imaging techniques that use $k$-space undersampling via different sampling schemes at acceleration factors have been developed to shorten scan durations. In this context, we propose a deep learning-based reconstruction method for 2D dynamic multi-contrast, multi-scheme, and multi-acceleration MRI. Our approach integrates the state-of-the-art vSHARP model, which utilizes half-quadratic variable splitting and ADMM optimization, with a Variational Network serving as an Auxiliary Refinement Network (ARN)  to better adapt to the diverse nature of MCCMRI data. Specifically, the subsampled $k$-space data is fed into the ARN, which produces an initial prediction for the denoising step  used by vSHARP. This, along with the subsampled $k$-space, is then used by vSHARP to generate high-quality 2D sequence predictions. Our method outperforms traditional reconstruction techniques and other vSHARP-based models.

\keywords{Deep Cardiac MRI Reconstruction \and MRI Reconstruction via vSHARP \and Multi-contrast MRI Reconstruction }
\end{abstract}
\section{Introduction}

Cardiac Magnetic Resonance Imaging (CMRI) is a cornerstone clinical imaging modality that provides detailed 4D (3D + dynamic) images of the heart, essential for assessing and evaluating cardiovascular health, structure, and function. Multi-contrast CMRI (MCCMRI) enhances this capability by acquiring multiple imaging sequences with varying contrast weightings, such as cardiac cine, mapping, tagging, phase-contrast, and dark-blood \cite{Wang2024CMRxRecon2024AM}. These sequences capture different anatomical views, including long-axis, short-axis, outflow tract, and aortic, offering a comprehensive assessment of the heart.

Despite the clinical importance of MCCMRI, it is often hindered by extensive acquisition times, which increases susceptibility to periodic or aperiodic body motion and causes patient discomfort. A viable solution is to reduce these lengthy acquisition times by subsampling the $k$-space during the MCCMRI acquisition. This approach accelerates the process by scanning fewer $k$-space data points than required by the Nyquist sampling theorem \cite{Nyquist1928CertainTI,shannon1949communication}.

This is typically achieved by applying a variety of subsampling patterns up to a specific acceleration factor. Common techniques include Cartesian trajectories such as equidistant sampling (each column sampled at a uniform distance from the next and previous), random-column sampling, variable density point sampling, pseudo-radial sampling, etc. \cite{retrospective,yiasemis2022deep}. While these patterns are often applied uniformly across all time frames of the dynamic acquisition for ease of implementation, distinct patterns can also be applied to each time frame \cite{Tsao2003ktBA}, exploiting potential cross-frame complementary information.

Although subsampling the $k$-space accelerates MCCMRI, it also degrades the quality of the reconstructed images, rendering them less useful for clinical practice. The Cardiac MRI Reconstruction (CMRxRecon) challenge \cite{lyu2024state} held at MICCAI in 2023 demonstrated that deep learning (DL)-based approaches represent the state-of-the-art (SOTA) compared to traditional methods.

In this work, we build upon the method that achieved one of the highest reconstruction quality submissions (2nd place in the challenge with minimal differences from the winning team) and demonstrated high reconstruction speed (more than $5\times$ faster than the winning team) \cite{yiasemis2023deep}. Specifically, the authors used the variable Splitting Half-quadratic ADMM algorithm for Reconstruction of inverse-Problems (vSHARP) \cite{Yiasemis2023vSHARPVS}, which was adapted for 2D dynamic CMRI reconstruction. They demonstrated that a unified model was able to quickly (2-15s) and effectively reconstruct 4D CMRI volumes of varying contrast (Cine, T1, T2), views (short or long-axis), and acceleration factors ($4\times$, $8\times$, $10\times$).

Building on vSHARP, we introduce an enhanced version with an Auxiliary Refinement Network (ARN) for multi-contrast, multi-view, multi-scheme, and multi-acceleration 2D dynamic reconstruction. Specifically, we employ a Variational Network \cite{varnet} as the ARN, which generates an initial state for the auxiliary variable in vSHARP's ADMM denoising step, using the subsampled $k$-space measurements as input to inform the reconstruction process of the specific type of input data. We outline 2D dynamic MRI reconstruction in \Section{sec2}, introduce our proposed methods in \Section{sec3}, present our experiments in \Section{sec4}, and conclude with a discussion of our results in \Section{sec5}.
\section{Background}
\label{sec:sec2}
\subsection{Accelerated Dynamic CMRI Reconstruction}

In the context of accelerated 2D dynamic CMRI reconstruction, the objective is to produce a 2D dynamic image $\vec{x}^{*} \in \mathbb{C}^{n \times n_f }$ from a 2D dynamic subsampled multi-coil $k$-space $\vec{y} \in \mathbb{C}^{n \times n_{c} \times n_f}$, where $n$ denotes the spatial dimension of each time frame, $n_c$ the number of acquisition coils, and $n_f$ the total number of time frames in the dynamic acquisition. This can be formulated as a solution to a regularized least-squares minimization problem:

\begin{equation}
    \vec{x}^{*} = \argmin_{\vec{x} \in \mathbb{C}^{n \times n_f }} \frac{1}{2} \, \left| \left| \mathcal{T}_{\vec{U}, \vec{S}}(\vec{x}) - \vec{y} \right| \right|_2^2 + \rho \, \mathcal{G}(\vec{x}),
\label{eq:optim}
\end{equation}
\noindent
where $\mathcal{T}_{\vec{U}, \vec{S}}$ denotes the forward operator, which transforms the 2D dynamic image $\vec{x} \in \mathbb{C}^{n \times n_f}$ into separate coil 2D dynamic images using sensitivity maps $\mat{S} \in \mathbb{C}^{n \times n_{c} \times n_f}$, transforms them to the $k$-space domain by applying the two-dimensional Fourier transform $\mathcal{F}$, and subsamples them using a binary subsampling pattern $\mat{U}$. Moreover, $\mathcal{G}: \mathbb{C}^{n \times n_f} \rightarrow \mathbb{R}$ represents an arbitrary functional assumed to impose prior reconstruction information \cite{https://doi.org/10.1002/mrm.21391}. \Equation{optim} can be solved numerically using iterative methods \cite{doi:10.1137/1021044} such as gradient descent or ADMM.

\subsection{Accelerated Dynamic CMRI Reconstruction with vSHARP}
\label{sec:subsec2.2}
Artificial intelligence-based methods address \eqref{eq:optim} using deep neural networks. Although many approaches exist for dynamic CMRI reconstruction \cite{Kstner2020,Oscanoa2023,SUN2024110218}, the 3D vSHARP \cite{yiasemis2023deep} forms the basis of our work due to its SOTA performance in CMRxRecon 2023. vSHARP approaches a solution to \eqref{eq:optim} by applying half-quadratic variable splitting (HQVS) followed by ADMM unrolling in $N$ iterations. Each iteration consists of three steps:
\begin{subequations}
\begin{equation}
    \vec{z}^{(j+1)} = \argmin_{\vec{z}\in\mathbb{C}^{n \times n_f }} \mathcal{G}(\vec{z}) + \frac{\rho}{2} \big | \big | \vec{x}^{(j)} - \vec{z} + \frac{\vec{u}^{(j)}}{\rho} \big | \big |_2^2 := \mathcal{D}_{\boldsymbol{\theta}_{j}}(\vec{x}^{(j)}, \vec{z}^{(j)}, \frac{\vec{u}^{(j)}}{\rho}),
\label{eq:admm_z}
\end{equation}
\begin{equation}
    \vec{x}^{(j+1)} = \argmin_{\vec{x}\in\mathbb{C}^{n \times n_f }} \frac{1}{2} \left|\left| \mathcal{T}_{\vec{U}, \vec{S}}(\vec{x}) - \Tilde{\vec{y}}^{k}\right|\right|_2^2 + \frac{\rho}{2} \big | \big | \vec{x} - \vec{z}^{(j+1)} + \frac{\vec{u}^{(j)}}{\rho} \big | \big |_2^2,
\label{eq:admm_x}
\end{equation}
\begin{equation}
    \vec{u}^{(j+1)} = \vec{u}^{(j)} + \rho (\vec{x}^{(j+1)} - \vec{z}^{(j+1)}), \quad j=0,\cdots, N-1.
\label{eq:admm_u}
\end{equation}
\label{eq:admm}
\end{subequations}
\noindent
As shown in \eqref{eq:admm}, vSHARP performs alternating optimization with respect to three variables: $\vec{z}$, the auxiliary variable derived from HQVS; $\vec{x}$, the image prediction; and $\vec{u}$, the Lagrange multipliers. At each iteration, \eqref{eq:admm_z} is estimated using a DL-based denoiser denoted as $\mathcal{D}_{\boldsymbol{\theta}{j}}$, specifically a 3D U-Net \cite{çiçek20163dunetlearningdense}, while \eqref{eq:admm_x} is solved numerically via a gradient descent unrolled scheme over $N_{\vec{x}}$ steps. Furthermore, vSHARP initializes $\vec{z}^{(0)}, \vec{x}^{(0)}$ with SENSE reconstruction \cite{pruessmann1999sense}, and for $\vec{u}^{(0)}$, it employs a DL-based module $\mathcal{I}_{\boldsymbol{\psi}}$ using $\vec{x}^{(0)}$:

\begin{equation}
    \vec{z}^{(0)} = \vec{x}^{(0)} = \sum_{k=1}^{n_c}{\mat{S}^{k}}^{*} \mathcal{F}^{-1} (\mat{y}^{k}), \quad \vec{u}^{(0)} = \mathcal{I}_{\boldsymbol{\psi}}( \vec{x}^{(0)}).
    \label{eq_init}
\end{equation}
\noindent
Consistent with \cite{yiasemis2023deep}, sensitivity maps are estimated using the autocalibration region of the subsampled data and refined via a 2D U-Net \cite{Ronneberger2015}.
\section{Methods}
\label{sec:sec3}
\subsection{Deep Learning Framework}
\label{sec:subsec3.1}
\subsubsection{Auxiliary Refinement Network}

To adapt to the diverse nature of our data, which vary in contrast, subsampling scheme, and acceleration factor, we incorporate an auxiliary refinement network (ARN), denoted as $\mathcal{A}_{\boldsymbol{\omega}}$. This network generates an initial image from the subsampled data, which serves as the initialization for the auxiliary variable in vSHARP's ADMM, replacing the SENSE reconstruction in \eqref{eq_init} for $\vec{z}^{(0)}$:

\begin{equation}
    \vec{z}^{(0)} = \sum_{k=1}^{n_c}{\mat{S}^{k}}^{*} \mathcal{F}^{-1} ( \vec{r}^{k} ), \quad \vec{r} = \text{DC}_{\mat{U}, \vec{y}} \circ \mathcal{A}_{\boldsymbol{\omega}} (\mat{y}, \mat{S}),
    \label{eq:init_arn}
\end{equation}
where $\mathcal{A}_{\boldsymbol{\omega}}$ can be any algorithm that inputs $k$-space data and complementary sensitivity maps, and outputs refined $k$-space data, and $\text{DC}$ denotes the data consistency operator (defined in \Section{subsec3.4}).  We opted for a 3D extension of the Variational Network \cite{varnet}, utilizing 3D U-Nets as regularizers with four scales and 12 channels in the first scale, and implemented 8 cascades.

\subsubsection{Reconstruction Model}
Our experiments utilize vSHARP as the reconstruction method, as detailed in \Section{subsec2.2}, with $N=12$ optimization iterations. We employ 3D U-Nets with four scales and 32 channels in the first scale for \eqref{eq:admm_z}, and perform $N_{\vec{x}}=6$ gradient descent steps for \eqref{eq:admm_x}. Sensitivity map estimation is performed using a 2D U-Net with four scales and 32 channels in the first scale.

\subsection{Implementation \& Training Details}
Our experiments, including model implementation, data handling, and subsampling, were conducted using the Deep Image Reconstruction Toolkit \cite{yiasemis2022direct}, on NVIDIA A100 or H100 GPUs.

\subsubsection{Optimization}

We trained our models using the Adam optimizer, starting with a learning rate of 1.6e-4, which was linearly increased to 5e-4 over the first 2k iterations, and subsequently decreased by 5\% every 50k iterations. Due to the large input $k$-space data size and the gradients computed during training, we utilized mixed precision training \cite{micikevicius2018mixedprecisiontraining} to allow for larger model implementations. To ensure training stability, we clipped all gradients' norms at 10.0 \cite{pmlr-v28-pascanu13}.

\subsubsection{Data Augmentations}

During training, we employed random augmentations. Specifically, we applied horizontal or vertical flips with a 50\% probability, as well as time-reversal with a 50\% probability, which involved reversing the time sequence. These augmentations were performed in the image domain by transforming the $k$-space data to the image domain, applying the augmentations, and then transforming it back to the $k$-space domain.

\subsection{Data Consistency}
\label{sec:subsec3.4}
Our reconstruction model, vSHARP, generates a series of image predictions from each ADMM optimization iteration $\{\vec{x}^{(j)}\}_{j=1}^{N}$. To ensure each prediction is consistent with the initial measurements $\vec{y}$, we enforce hard data consistency:

\begin{equation}
    \hat{\vec{x}}^{(j)} = \sum_{k=1}^{n_c}{\mat{S}^{k}}^{*} \mathcal{F}^{-1} ((\hat{\mat{y}}^{(j)})^{k}), \quad  \hat{\mat{y}}^{(j)} = \text{DC}_{\mat{U}, \vec{y}}\big( \mat{S}^{1}\mathcal{F}(\vec{x}^{(j)}), \cdots, \mat{S}^{n_c}\mathcal{F}(\vec{x}^{(j)}) \big),
\end{equation}
where $\text{DC}_{\mat{M}}$ is the hard data consistency operator defined as:
\begin{equation}
    \text{DC}_{\mat{U}, \vec{y}}(\vec{w}) = \mat{U}(\vec{y}) + (\mathbb{1} - \mat{U})(\vec{w}).
\end{equation}

\subsection{Loss Function}

Given that our training data is fully sampled, we have access to ground truth images and $k$-space data, enabling us to train our models in a supervised manner using a dual-domain loss strategy. This involves computing the loss in both the image and frequency domains. Specifically, assuming $\vec{x}^{*}$ and $\vec{y}^{*}$ are the ground truth root-sum-of-squares image and fully sampled $k$-space, we define the following loss function:
\begin{equation}
\begin{gathered}
     \sum_{t=1}^{n_f} \Big[ \mathcal{L}_\text{SSIM}(|\vec{z}_t^{(0)}|, \vec{x}_t^*) + \big|\big| |\vec{z}_t^{(0)}| -  \vec{x}_t^* \big|\big|_1  \Big] 
     + \frac{|| \vec{r} - \vec{y}^{*}||_1}{|| \vec{y}^{*}|||_1}  + \mathcal{L}_\text{SSIM3D}(|\vec{z}^{(0)}|, \vec{x}^*)  \\
    + \sum_{j=1}^{N} 10^\frac{(j-N)}{(N-1)} \Big( \sum_{t=1}^{n_f} \Big[ \mathcal{L}_\text{SSIM}(|\hat{\vec{x}}_t^{(j)}|, \vec{x}_t^*) + \big|\big| |\hat{\vec{x}}_t^{(j)}| -  \vec{x}_t^* \big|\big|_1  \Big] \\
     + \frac{|| \hat{\vec{y}}^{(j)} - \vec{y}^{*}||_1}{|| \vec{y}^{*}|||_1}  + \mathcal{L}_\text{SSIM3D}(|\hat{\vec{x}}^{(j)}|, \vec{x}^*)\Big),
\end{gathered}
\end{equation}
\normalsize
where $\vec{z}^{(0)}$ and $\vec{r}$ as in \Section{subsec3.1}, and $\{\hat{\vec{x}}^{(j)}\}_{j=1}^{N}$ and $\{\hat{\vec{y}}^{(j)}\}_{j=1}^{N}$ are as defined in \Section{subsec3.4}. $\mathcal{L}_\text{SSIM}$ and $ \mathcal{L}_\text{SSIM3D}$ are as defined in the literature \cite{yiasemis2023deep}.

\subsection{Evaluation Metrics}
For evaluation, we employed three common image reconstruction fidelity metrics: structural similarity index measure (SSIM), peak-signal-to-noise ratio (PSNR), and normalized mean squared error (NMSE).

\section{Experiments}
\label{sec:sec4}

\subsection{Data}
The dataset used for our experiments consists of 1,404 multi-contrast 5D (3D + time + coils) volumes of $k$-space data, including cardiac cine, T1/T2 mapping, tagging, phase-contrast (flow2d), and dark-blood imaging \cite{Wang2024CMRxRecon2024AM}. These were acquired using a 3T MRI scanner with protocols outlined in previous studies \cite{Wang2021}. The data encompasses various anatomical views, such as long-axis (LAX) views (2, 3, and 4-chamber), short-axis (SAX), left ventricular outflow tract (LVOT), and aortic (transversal and sagittal) views. The training set includes fully-sampled cine, aorta, mapping, and tagging data, while the inference set includes subsampled data and two additional unseen contrasts (flow2d and black-blood). 

\subsection{Experimental Setups}

We consider two different experimental setups:
\begin{enumerate}[label=(\textbf{\Alph*}),labelindent=0pt]
    \item Multi-contrast with equispaced (same mask applied to all time-frames) sampling at accelerations  $R=4,8,10$, evaluated on seen and unseen modalities (1,614 volumes in total).
    \item Multi-contrast with $k$t (interleaved masks applied across time-frames) multi-scheme (equispaced, Gaussian 1D, radial) sampling, at $R=4-24$, evaluated on seen modalities (1,281 volumes in total).
\end{enumerate}
\noindent
For each experimental setup, we trained separate models responsible for reconstructing inference data within each setup, accommodating any contrast, view, acceleration, and subsampling scheme (applicable only for \textbf{B}).

\subsection{Comparisons}
We compared the following methods:
\begin{enumerate}
    \item Zero-filled reconstruction (no optimization).
    \item Two traditional reconstruction algorithms: GRAPPA \cite{https://doi.org/10.1002/mrm.10171}, SENSE \cite{pruessmann1999sense}.
    \item Proposed method (vSHARP without ARN module as outlined in \Section{subsec3.1}).
    \item Variants of the proposed method:
    \begin{enumerate}
        \item vSHARP with an ARN configuration with 8 channels in the first scale of each regularizer, denoted as (S).
        \item vSHARP with $N=14$, $N_\vec{x}=8$, denoted as (L) with an ARN as outlined in \Section{subsec3.1}. For training this model, $k$-space “cropping” augmentations following \cite{yiasemis2023deep} were performed to regulate memory during training.
    \end{enumerate}
    Note that for these methods in experimental setup \textbf{A}, distinct models for each acceleration factor were trained.
\end{enumerate}

\subsection{Results}
\begin{figure}[!htb]
    \centering
    \includegraphics[width=1\textwidth]{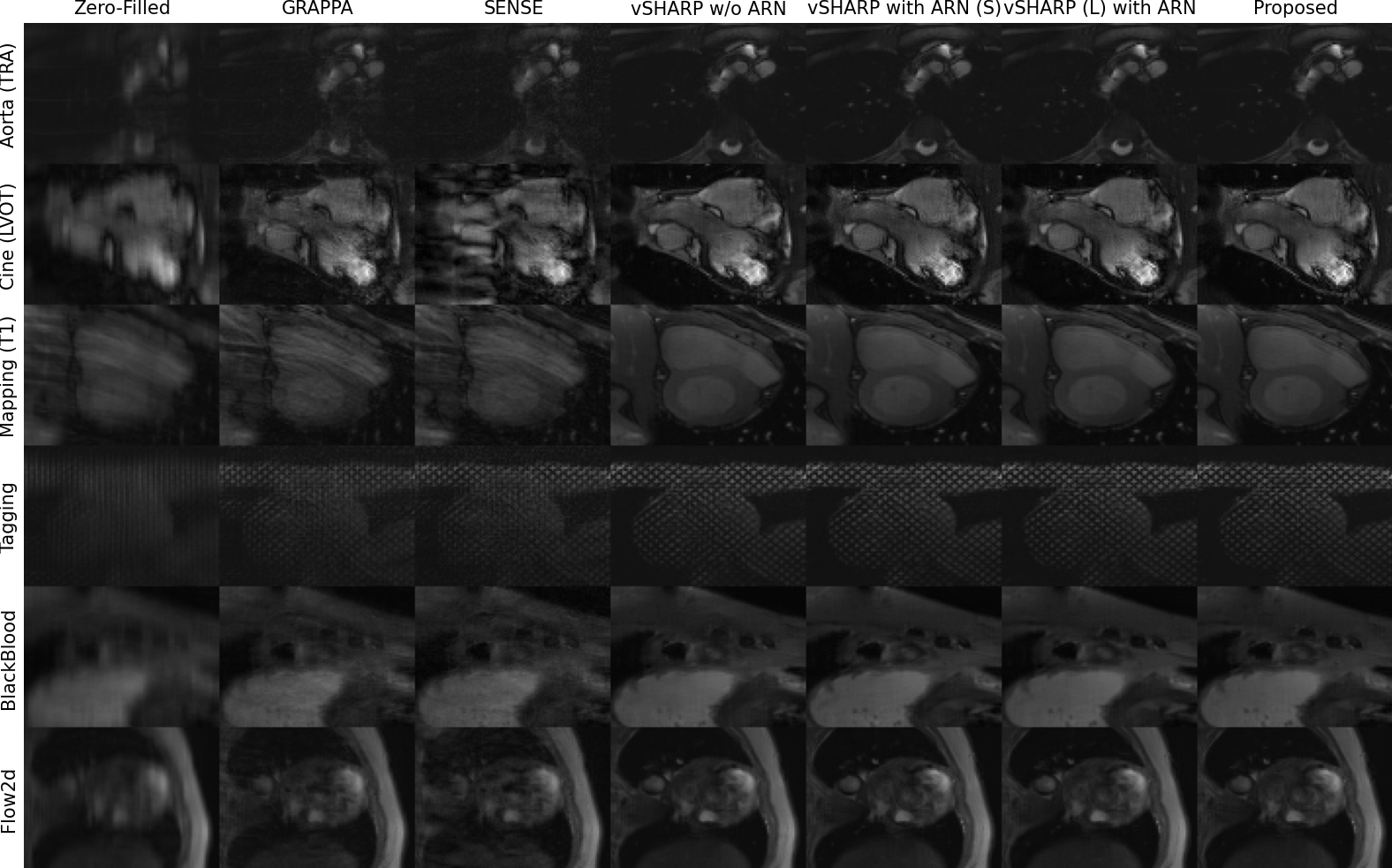}
    \caption{Sample reconstructions from the inference set in Experimental Setup \textbf{A}.}
    \label{fig:recons_A}
\end{figure}

Tables \ref{tab:metrics_uni} and \ref{tab:metrics_$k$t} summarize the results of our experiments for setups $\mathbf{A}$ and $\mathbf{B}$, respectively. Sample reconstructions at $10\times$ acceleration from setup $\textbf{A}$ are illustrated in \Figure{recons_A}. We also present the average inference times for reconstructing a 5D $k$-space volume using both the original vSHARP and our proposed method in \Table{inference_times} acquired on the same machine (NVIDIA A100). 


\setlength{\tabcolsep}{10.0pt}
\renewcommand{\arraystretch}{1.2}
\begin{table}[!hbt]
\centering
\caption{Experimental setup $\mathbf{A}$ results using different methods. Metrics are averaged across contrasts and acceleration factors.}
\label{tab:metrics_uni}
\resizebox{\textwidth}{!}{%
\begin{tabular}{ccccccc}
\hline 
\multirow{2}{*}{\textbf{Method}} &  & \multicolumn{5}{c}{\textbf{Average Metrics}}               \\ \cline{3-7} 
                                 &  & \textbf{SSIM}   &  & \textbf{PSNR}    &  & \textbf{NMSE}   \\ \cline{1-1} \cline{3-3} \cline{5-5} \cline{7-7}
Zero-Filled                      &  & 0.5026          &  & 20.5536          &  & 0.238           \\
GRAPPA                           &  & 0.7351          &  & 28.3721          &  & 0.0921          \\
SENSE                            &  & 0.7041          &  & 27.1099          &  & 0.1125          \\
vSHARP w/o ARN                   &  & 0.8789          &  & 32.7229          &  & 0.0321          \\
vSHARP with ARN (S)              &  & 0.8937          &  & 33.5504          &  & 0.0259          \\
vSHARP (L) with ARN              &  & 0.8911          &  & 33.4788          &  & 0.0264          \\
vSHARP with ARN (proposed)       &  & \textbf{0.9014} &  & \textbf{33.9011} &  & \textbf{0.0242} \\ \hline 
\end{tabular}%
}
\end{table}


\setlength{\tabcolsep}{10.0pt}
\renewcommand{\arraystretch}{1.2}
\begin{table}[!hbt]
\centering
\caption{Experimental setup $\mathbf{B}$ results using different methods. Metrics are averaged across contrasts and acceleration factors.}
\label{tab:metrics_$k$t}
\resizebox{\textwidth}{!}{%
\begin{tabular}{ccccccc}
\hline
\multirow{2}{*}{\textbf{Method}} &           & \multicolumn{5}{c}{\textbf{Average Metrics}}                        \\ \cline{3-7} 
                                 & \textbf{} & \textbf{SSIM}   & \textbf{} & \textbf{PSNR}    &  & \textbf{NMSE}   \\ \cline{1-1} \cline{3-3} \cline{5-5} \cline{7-7} 
Zero-Filled                      &           & 0.4610          &           & 19.4182          &  & 0.2853          \\
GRAPPA                           &           & 0.6449          &           & 24.9678          &  & 0.1554          \\
SENSE                            &           & 0.6200          &           & 24.4780          &  & 0.1734          \\
vSHARP w/o ARN                   &           & 0.8421          &           & 30.6929          &  & 0.0428          \\
vSHARP with ARN (S)              &           & 0.8639          &           & 31.6883          &  & 0.0408          \\
vSHARP (L) with ARN              &           & 0.8555          &           & 31.3105          &  & 0.0448          \\
vSHARP with ARN (proposed)       &           & \textbf{0.8810} &           & \textbf{32.4700} &  & \textbf{0.0346} \\ \hline
\end{tabular}%
}
\end{table}

\setlength{\tabcolsep}{10.0pt}
\renewcommand{\arraystretch}{1.2}
\begin{table}[!hbt]
\centering
\caption{Average inference time per 5D volume (in seconds).}
\label{tab:inference_times}
\resizebox{1\textwidth}{!}{%
\begin{tabular}{lcc}
\hline
\multirow{2}{*}{\textbf{Experimental Setup}} & \multicolumn{2}{c}{\textbf{Inference Time (s)}} \\ \cline{2-3} 
                                 & vSHARP w/o ARN & vSHARP with ARN (proposed) \\ \hline
\multicolumn{1}{c}{$\textbf{A}$} & 7.06           & 7.78                       \\
\multicolumn{1}{c}{$\textbf{B}$} & 9.01           & 10.40                      \\ \hline
\end{tabular}%
}
\end{table}

As shown from the results, our proposed method consistently outperformed all other techniques across both experimental setups. It achieved the best scores in all metrics, including SSIM, PSNR, and NMSE, surpassing traditional methods such as GRAPPA and SENSE, as well as other variations of the vSHARP method. Interestingly, the models trained for individual acceleration factors in setup \textbf{A} did not surpass our unified model trained across all acceleration factors. In addition, Figure \ref{fig:recons_A} shows that vSHARP based reconstructions are superior compared to traditional methods. Also, we can see that all vSHARP variants are able to reconstruct well the two unseen modalities.

In terms of reconstruction times, as expected our proposed method shows an increase of around 0.7 to 1.4 seconds, as evident from \Table{inference_times}.

\section{Discussion and Conclusion}
\label{sec:sec5}
This study introduces a new deep learning-based reconstruction method that enhances the vSHARP algorithm with an Auxiliary Refinement Network (ARN) for accelerated multi-contrast cardiac MRI. Integrating the ARN into the vSHARP framework improves reconstruction quality, outperforming traditional methods and other vSHARP variants. The method's effectiveness across a wide range of imaging sequences, views, and acceleration factors demonstrates its practical utility in clinical settings, where the use of multiple specialized models may not be feasible. However, it should be noted that the use of ARN introduces an increase in reconstruction time.

Our experiments focused exclusively on a single type of ARN—the Variational Network, a deep learning-based approach.  Alternative ARNs, including traditional techniques like GRAPPA or SENSE, or other deep learning-based methods, could be investigated, though non-DL approaches may require significantly more computation time. Additionally, we employed the ARN to inform the reconstruction model about input types (contrast, view, subsampling scheme, acceleration factor) via auxiliary variable initialization. Other conditioning methods, such as using a multi-class label input to a trained embedding space model, could also be explored.

Future work should focus on incorporating additional contrasts, subsampling schemes, and anatomical regions into the training data to further improve the model's robustness and generalizability.
%
%

%
%
\bibliographystyle{splncs04}
\bibliography{bib}

\begin{thebibliography}{10}
\providecommand{\url}[1]{\texttt{#1}}
\providecommand{\urlprefix}{URL }
\providecommand{\doi}[1]{https://doi.org/#1}

\bibitem{https://doi.org/10.1002/mrm.10171}
Griswold, M.A., Jakob, P.M., Heidemann, R.M., Nittka, M., Jellus, V., Wang, J., Kiefer, B., Haase, A.: Generalized autocalibrating partially parallel acquisitions (grappa). Magnetic Resonance in Medicine  \textbf{47}(6),  1202--1210 (2002). \doi{https://doi.org/10.1002/mrm.10171}, \url{https://onlinelibrary.wiley.com/doi/abs/10.1002/mrm.10171}

\bibitem{Kstner2020}
K\"{u}stner, T., Fuin, N., Hammernik, K., Bustin, A., Qi, H., Hajhosseiny, R., Masci, P.G., Neji, R., Rueckert, D., Botnar, R.M., Prieto, C.: Cinenet: deep learning-based 3d cardiac cine mri reconstruction with multi-coil complex-valued 4d spatio-temporal convolutions. Scientific Reports  \textbf{10}(1) (Aug 2020). \doi{10.1038/s41598-020-70551-8}, \url{http://dx.doi.org/10.1038/s41598-020-70551-8}

\bibitem{https://doi.org/10.1002/mrm.21391}
Lustig, M., Donoho, D., Pauly, J.M.: Sparse mri: The application of compressed sensing for rapid mr imaging. Magnetic Resonance in Medicine  \textbf{58}(6),  1182--1195 (2007). \doi{https://doi.org/10.1002/mrm.21391}, \url{https://onlinelibrary.wiley.com/doi/abs/10.1002/mrm.21391}

\bibitem{lyu2024state}
Lyu, J., Qin, C., Wang, S., Wang, F., Li, Y., Wang, Z., Guo, K., Ouyang, C., T{\"a}nzer, M., Liu, M., et~al.: The state-of-the-art in cardiac mri reconstruction: Results of the cmrxrecon challenge in miccai 2023. arXiv preprint arXiv:2404.01082  (2024)

\bibitem{micikevicius2018mixedprecisiontraining}
Micikevicius, P., Narang, S., Alben, J., Diamos, G., Elsen, E., Garcia, D., Ginsburg, B., Houston, M., Kuchaiev, O., Venkatesh, G., Wu, H.: Mixed precision training (2018), \url{https://arxiv.org/abs/1710.03740}

\bibitem{Nyquist1928CertainTI}
Nyquist, H.: Certain topics in telegraph transmission theory. Transactions of the American Institute of Electrical Engineers  \textbf{47},  617--644 (1928), \url{https://api.semanticscholar.org/CorpusID:8632488}

\bibitem{Oscanoa2023}
Oscanoa, J.A., Middione, M.J., Alkan, C., Yurt, M., Loecher, M., Vasanawala, S.S., Ennis, D.B.: Deep learning-based reconstruction for cardiac mri: A review. Bioengineering  \textbf{10}(3), ~334 (Mar 2023). \doi{10.3390/bioengineering10030334}, \url{http://dx.doi.org/10.3390/bioengineering10030334}

\bibitem{pmlr-v28-pascanu13}
Pascanu, R., Mikolov, T., Bengio, Y.: On the difficulty of training recurrent neural networks. In: Dasgupta, S., McAllester, D. (eds.) Proceedings of the 30th International Conference on Machine Learning. Proceedings of Machine Learning Research, vol.~28, pp. 1310--1318. PMLR, Atlanta, Georgia, USA (17--19 Jun 2013), \url{https://proceedings.mlr.press/v28/pascanu13.html}

\bibitem{pruessmann1999sense}
Pruessmann, K.P., Weiger, M., Scheidegger, M.B., Boesiger, P.: Sense: sensitivity encoding for fast mri. Magnetic Resonance in Medicine: An Official Journal of the International Society for Magnetic Resonance in Medicine  \textbf{42}(5),  952--962 (1999)

\bibitem{Ronneberger2015}
Ronneberger, O., Fischer, P., Brox, T.: U-net: Convolutional networks for biomedical image segmentation. In: Lecture Notes in Computer Science, pp. 234--241. Springer International Publishing (2015). \doi{10.1007/978-3-319-24574-4\_28}, \url{https://doi.org/10.1007/978-3-319-24574-4\_28}

\bibitem{shannon1949communication}
Shannon, C.E.: Communication in the presence of noise. Proceedings of the IRE  \textbf{37}(1),  10--21 (1949)

\bibitem{varnet}
Sriram, A., Zbontar, J., Murrell, T., Defazio, A., Zitnick, C.L., Yakubova, N., Knoll, F., Johnson, P.: End-to-end variational networks for accelerated mri reconstruction. In: Martel, A.L., Abolmaesumi, P., Stoyanov, D., Mateus, D., Zuluaga, M.A., Zhou, S.K., Racoceanu, D., Joskowicz, L. (eds.) Medical Image Computing and Computer Assisted Intervention -- MICCAI 2020. pp. 64--73. Springer International Publishing, Cham (2020)

\bibitem{SUN2024110218}
Sun, J., Wang, C., Guo, L., Fang, Y., Huang, J., Qiu, B.: An unrolled neural network for accelerated dynamic mri based on second-order half-quadratic splitting model. Magnetic Resonance Imaging  \textbf{113},  110218 (2024). \doi{https://doi.org/10.1016/j.mri.2024.110218}, \url{https://www.sciencedirect.com/science/article/pii/S0730725X24001930}

\bibitem{Tsao2003ktBA}
Tsao, J., Boesiger, P., Pruessmann, K.P.: k‐t blast and k‐t sense: Dynamic mri with high frame rate exploiting spatiotemporal correlations. Magnetic Resonance in Medicine  \textbf{50} (2003), \url{https://api.semanticscholar.org/CorpusID:9806611}

\bibitem{Wang2021}
Wang, C., Li, Y., Lv, J., Jin, J., Hu, X., Kuang, X., Chen, W., Wang, H.: Recommendation for cardiac magnetic resonance imaging-based phenotypic study: Imaging part. Phenomics  \textbf{1}(4),  151–170 (Jul 2021). \doi{10.1007/s43657-021-00018-x}, \url{http://dx.doi.org/10.1007/s43657-021-00018-x}

\bibitem{Wang2024CMRxRecon2024AM}
Wang, Z., Wang, F., Qin, C., Lyu, J., Cheng, O., Wang, S., Li, Y., Yu, M., Zhang, H., Guo, K., Shi, Z., Li, Q., Xu, Z., Zhang, Y., Li, H., Hua, S., Chen, B., Sun, L., qi~Sun, M., Li, Q., Chu, Y., Bai, W., Qin, J., Zhuang, X., Prieto, C., Young, A., Markl, M., Wang, H., pan Wu, L., Yang, G., Qu, X., Wang, C.: Cmrxrecon2024: A multi-modality, multi-view k-space dataset boosting universal machine learning for accelerated cardiac mri. ArXiv  \textbf{abs/2406.19043} (2024)

\bibitem{doi:10.1137/1021044}
Willoughby, R.A.: Solutions of ill-posed problems (a. n. tikhonov and v. y. arsenin). SIAM Review  \textbf{21}(2),  266--267 (1979). \doi{10.1137/1021044}, \url{https://doi.org/10.1137/1021044}

\bibitem{yiasemis2022direct}
Yiasemis, G., Moriakov, N., Karkalousos, D., Caan, M., Teuwen, J.: Direct: Deep image reconstruction toolkit. Journal of Open Source Software  \textbf{7}(73), ~4278 (2022)

\bibitem{yiasemis2023deep}
Yiasemis, G., Moriakov, N., Sonke, J.J., Teuwen, J.: Deep cardiac mri reconstruction with admm. In: International Workshop on Statistical Atlases and Computational Models of the Heart. pp. 479--490. Springer (2023)

\bibitem{Yiasemis2023vSHARPVS}
Yiasemis, G., Moriakov, N., Sonke, J.J., Teuwen, J.: vsharp: variable splitting half-quadratic admm algorithm for reconstruction of inverse-problems. Magnetic Resonance Imaging p. 110266 (2024)

\bibitem{retrospective}
Yiasemis, G., S{\'a}nchez, C., Sonke, J.J., Teuwen, J.: On retrospective k-space subsampling schemes for deep mri reconstruction. Magnetic Resonance Imaging  \textbf{107},  33--46 (Apr 2024). \doi{10.1016/j.mri.2023.12.012}

\bibitem{yiasemis2022deep}
Yiasemis, G., Zhang, C., S{\'a}nchez, C.I., Sonke, J.J., Teuwen, J.: Deep mri reconstruction with radial subsampling. In: Medical Imaging 2022: Physics of Medical Imaging. vol. 12031, pp. 801--810. SPIE (2022)

\bibitem{çiçek20163dunetlearningdense}
Özgün Çiçek, Abdulkadir, A., Lienkamp, S.S., Brox, T., Ronneberger, O.: 3d u-net: Learning dense volumetric segmentation from sparse annotation (2016), \url{https://arxiv.org/abs/1606.06650}

\end{thebibliography}

\end{document}